\documentclass[11pt,a4paper]{article}
\usepackage[utf8]{inputenc}
\usepackage{amsmath,amssymb,cite,epsfig}
\usepackage{xcolor}
\usepackage{graphicx}
\usepackage{hyperref}

\begin{document}

\title{Stability of spherical thin-shell wormholes in scalar-tensor theories} 
\author{Ernesto F. Eiroa$^{1}$\thanks{e-mail: eiroa@iafe.uba.ar}, Griselda Figueroa-Aguirre$^{1}$\thanks{e-mail: gfigueroa@iafe.uba.ar}, Vasiliki Karanasou$^{2}$\thanks{e-mail: vasiliki.karanasou@ut.ee}\\
{\small $^1$ Instituto de Astronom\'{\i}a y F\'{\i}sica del Espacio (IAFE, CONICET-UBA),}\\
{\small Ciudad Universitaria, 1428, Buenos Aires, Argentina}\\
{\small $^2$ Laboratory of Theoretical Physics, Institute of Physics,} \\ 
{\small University of Tartu, W. Ostwaldi 1, 50411 Tartu, Estonia}} 

\maketitle

\begin{abstract}

In this article, we construct a family of spherically symmetric thin-shell wormholes within scalar-tensor theories of gravity. In the case of wormholes symmetric across the throat, we study the matter content and analyze the stability of the static configurations under radial perturbations. We apply the formalism to a particular example involving Einstein-Maxwell gravity coupled to a conformally invariant scalar field. We show that stable configurations are possible for suitable values of the parameters involved.

\end{abstract}

\section{Introduction}\label{intro} 

General Relativity (GR) is a very successful theory of gravity, which has provided explanations for a variety of astrophysical and cosmological phenomena, and has passed many experimental tests \cite{Will:2014kxa, Baker:2014zba, yunes:2025}. However, there are some unresolved issues, such as the elusive nature of the dark matter \cite{Bertone:2018,Oks:2021} necessary to model various observations, the cosmological data indicating that the late-time expansion of the Universe is accelerated \cite{SupernovaCosmologyProject:1998vns, SupernovaSearchTeam:1998fmf}, which requires the presence of dark energy \cite{Oks:2021}, and the so-called Hubble tension \cite{DiValentino:2021izs, Schoneberg:2021qvd}. All of these suggest that GR might not be the final answer. In this context, several theories of modified  gravity have appeared in the literature. Among them, the scalar-tensor theories are interesting alternatives to GR, in which  gravity is mediated by a tensor and a scalar field \cite{Fujii_Maeda_2003}. They are generalizations of the Brans-Dicke theory \cite{Brans:1961sx}. Several topics have been studied within scalar-tensor gravity, such as gravitational waves \cite{Wagoner:1970vr}, the post-Newtonian expansion \cite{Nordtvedt:1970uv}, black holes \cite{Mart_nez_2003, Mart_nez_2006, Dotti:2008, Astorino:2013sfa}, and cosmological aspects \cite{Faraoni:2004pi}. It is worth mentioning the existence of an ambiguity in scalar-tensor theories related to the distinction between the Einstein and the Jordan frames. There is a long debate about the most suitable physical frame and the (non)equivalence of them \cite{Faraoni:1999hp, flanagan:2004}. Another proposal is $F(R)$ gravity \cite{nojiri:2011, nojiri:2017}, in which the Lagrangian density is a function of the Ricci scalar $R$. More alternative theories to GR can be found in \cite{nojiri:2017}.

In the context of the different theories of gravity, there are solutions of the field equations that represent compact objects, such as black holes and wormholes. The existence of wormholes is still hypothetical; they were first introduced by Einstein and  Rosen \cite{Einstein:1935tc} and studied more systematically since the work by Morris and Thorne \cite{Morris:1988cz}. These geometries represent bridges that connect two different regions of the Universe or two different universes across a throat \cite{Visser:1995cc}, where the flare-out condition is fulfilled. A common issue encountered in traversable wormholes is the presence of the so-called exotic matter, i.e. matter that does not satisfy the energy conditions  \cite{Visser:1995cc}, which is not a physically desirable feature. In GR, the exotic matter is always required near the throat \cite{Hochberg:1997, Hochberg:1998ii, Hochberg:1998ha}; however, its amount can be reduced \cite{Visser:2003}, but at the cost of increasing tensions in it \cite{Zaslavskii:2007}. Many papers concerning wormholes with scalar fields have appeared in the literature, here we can only mention a few \cite{Kim:1997jf, Barcelo:1999hq, Barcelo:2000zf, Bronnikov:2005an, Bronnikov:2010tt, Butcher:2015sea, Hohmann:2018shl, Papantonopoulos:2019ugr, DeFalco:2021ksd, Bronnikov:2022bud, Nojiri:2023dvf, Cadoni:2025cmf}. On the other hand, a new spacetime can be obtained by the cut and paste of two geometries, which are joined on their boundaries by applying the suitable junction conditions, firstly found in GR \cite{Darmois:1927, Israel:1966rt}. This method is used to construct the denominated thin-shell wormholes \cite{Visser:1995cc}, in which the new manifold has an infinitesimally thin layer of matter at the joining hypersurface, where the throat is placed. This procedure allows to localize and minimize the exotic matter at the throat. Although matter not satisfying the energy conditions appears in many physical contexts, in the case of wormholes is advantageous to reduce the amount required at the throat. Another interesting aspect of thin-shell wormholes is the possibility of performing analytical studies of stability in the case of highly symmetric static configurations, under perturbations preserving the symmetry. The construction, the stability, and other related topics on thin-shell wormholes have been considered within GR \cite{Poisson:1995sv, Eiroa:2003wp, Lobo:2003xd, Eiroa:2008ky, PhysRevD.86.044026, PhysRevD.92.044002, Sharif:2021}, in dilaton gravity \cite{Eiroa:2005pc}, in Einstein-Gauss-Bonnet \cite{Thibeault:2005ha, Richarte:2007zz}, in Brans-Dicke \cite{Eiroa:2008hv}, in $f(R)$ theory \cite{Sharif_2014, Eiroa:2015hrt, Eiroa:2016zjx, Mazharimousavi:2020rra, Eiroa:2020abc, Godani:2023tcx, Rosa:2023olc}, and also in Palatini $f(R)$ gravity \cite{Lobo:2020vqh}, among others.
 
In this paper, we construct thin-shell wormholes with spherical symmetry in Einstein theory non-minimally coupled to a scalar field. In order to have a proper matching at the shell, we apply the junction conditions corresponding to this theory \cite{Sakai:1992ud, Aviles:2019xae}. For a family of wormholes symmetric across the throat, which is located at the shell, we study their dynamical stability and the presence of exotic matter. In particular, for our construction we adopt a black hole solution within Einstein-Maxwell gravity coupled to a conformally invariant scalar field, having a constant scalar field  \cite{Astorino:2013sfa}. The structure of this paper is as follows. In Sect. \ref{tswh}, we present the general construction of the wormholes in scalar-tensor gravity. In Sect. \ref{stab}, we develop the formalism for the stability of wormholes symmetric across the throat. In Sect. \ref{scalartensor}, we consider a particular example of a spacetime with a radial electric field, in which we study the stability and the energy conditions. Finally, we summarize our results in Sect. \ref{conclu}. We use units such that $c=1$.

\section{Wormhole construction}\label{tswh}

In this section, we construct a class of spherically symmetric thin-shell wormholes in scalar-tensor gravity by adopting the so-called junction formalism. We are interested in a class of scalar-tensor theories with a non-minimally coupled real scalar field $\phi$,  a self-interaction potential $U(\phi)$, and an arbitrary function $f(\phi )$, described by the action in the Jordan frame 
\begin{equation}
S = \int d^4x \, \sqrt{-g} \left(f(\phi) R 
- \frac{1}{2}\nabla_{\mu}\phi\nabla^{\mu}\phi
- U(\phi) \right)+\int d^4x \, \sqrt{-g} \mathcal{L}_{M},
\label{action1}
\end{equation}
from which the following field equations are obtained
\begin{equation}
2f(\phi) G_{\mu\nu}+g_{\mu\nu}\left(\frac{1}{2}(\nabla\phi)^2+U(\phi)\right)-\nabla_{\mu}\phi\nabla_{\nu}\phi-2\nabla_{\mu}\nabla_{\nu}f(\phi)+2g_{\mu\nu} \Box f(\phi)=T_{\mu\nu},
\end{equation}
\begin{equation}
\Box \phi + f'(\phi)R-U'(\phi)=0,
\end{equation}
where the prime on the functions $f(\phi)$ and $U(\phi)$ represents the derivative with respect to $\phi$, and $T_{\mu\nu}$ is the energy-momentum tensor associated with the other matter field Lagrangian $\mathcal{L}_{M}$.

The junction formalism for these theories, introduced in Ref. \cite{Aviles:2019xae}, establishes the conditions for proper matching of two manifolds $\mathcal{M}^{1}$ and  $\mathcal{M}^{2}$ at a hypersurface $\Sigma$. We denote the jump of any quantity $\Upsilon  $ across $\Sigma$ by
\begin{equation}
[\Upsilon ]\equiv (\Upsilon ^{2}-\Upsilon  ^{1})|_\Sigma.
\end{equation}
The first set of conditions arises from the continuity of the metric and the scalar field across the matching hypersurface, respectively
\begin{equation}
[g_{\mu \nu}]=0,
\label{jc1}
\end{equation}
\begin{equation}
[\phi]=0 ,
\label{jc2}
\end{equation}
while the second set comes from the Einstein equations and the equation of motion of the scalar field, respectively
\begin{equation}
-2\varepsilon f(\phi) ([K_{\mu \nu}]-h_{\mu \nu}[K]) + 2\Omega f'(\phi)h_{\mu\nu} = S_{\mu\nu} ,
\label{jc3}
\end{equation}
\begin{equation}
\Omega = 2 \varepsilon f'(\phi)[K].
\label{jc4}
\end{equation}
In these equations, $h_{\mu\nu}$ is the first fundamental form, $K_{\mu\nu}$ is the second fundamental form, $K=K^{\mu}_{\;\; \mu}$ is its trace, $S_{\mu \nu}$ is the energy-momentum tensor at $\Sigma$, and $\Omega$ is defined by
\begin{equation}
\Omega = \varepsilon [n^{\mu} \partial_{\mu}\phi],
\label{Mdef}
\end{equation}
where $n^{\gamma}$ is the unit normal to $\Sigma$ that points from $\mathcal{M}^{1}$ to  $\mathcal{M}^{2}$, satisfying $n^{\gamma }n_{\gamma }=\varepsilon$. The hypersurface $\Sigma$ in our case will be timelike and so we set $\varepsilon=1$ from now on (the value $\varepsilon =-1$ corresponds to a spacelike hypersurface). One speaks of a boundary hypersurface when $S_{\mu \nu }\equiv 0$, otherwise, of a thin shell of matter.

In our construction, we start from two spherically symmetric spacetimes described by the line elements
\begin{equation} 
ds_{1,2}^2=-A_{1,2} (r) dt_{1,2}^2+A_{1,2} (r)^{-1} dr^2+r^2(d\theta^2 + \sin^2\theta d\varphi^2),
\label{metric}
\end{equation}
where $r\ge 0$ is the radial coordinate, while $0\le \theta \le \pi$ and $0\le \varphi<2\pi $ are the angular coordinates. We select a radius $a>0$ and 
remove the region with $r< a$ from each geometry, in order to define the manifolds
\begin{equation} 
\mathcal{M}^{1,2 }=\{X_{1,2}^{\alpha }=(t_{1,2},r,\theta,\varphi)/r\geq a\},  
\end{equation}
and we join them at the surface
\begin{equation} 
\Sigma \equiv \Sigma ^{1,2 }=\{X_{1,2}/\mathcal{G}(r)=r-a=0\}.
\end{equation}
This cut-and-paste procedure has as a result a new geodesically complete manifold $\mathcal{M}=\mathcal{M}^{1} \cup \mathcal{M}^{2}$. The area $4\pi r^2$ is minimal at $r=a$, so the flare-out condition is satisfied there and $\mathcal{M}$ describes a wormhole with two regions connected by the throat $\Sigma$ with radius $a$. We can use the proper radial distance $\ell =\pm \int_{a}^{r}\sqrt{1/A_{1,2} (r)}dr$ in order to introduce a global radial coordinate in $\mathcal{M}$, where the ($-$) and ($+$) sign refers to  $\mathcal{M}^{1}$ and to $\mathcal{M}^{2}$ respectively, with the wormhole throat placed at $\ell = 0$. 

Apart from the coordinates that describe the 4-dimensional spacetime, we introduce the coordinates $\xi ^{i}=(\tau ,\theta,\varphi )$ on the surface $\Sigma $, with $\tau $ the proper time on the shell. We let the throat radius $a$ to be a function of $\tau$   in order to study of the stability of the shell under radial perturbations. The line element should be continuous across $\Sigma $, then the time coordinates at each side satisfy that $d\tau^2 = A_1(a)^2\left( A_1(a) + \dot{a}^2\right) ^{-1}dt_1^2=A_2(a)^2\left( A_2(a) + \dot{a}^2\right) ^{-1}dt_2^2$, where the dot represents the derivative with respect to $\tau$. The induced metric on $\Sigma$ then reads
\begin{equation}
ds_\Sigma^2=-d\tau^2+a^2(\tau)(d\theta ^2+\sin^2\theta d\varphi ^2).
\end{equation}
The first fundamental form $h_{ij}$ (the metric on the shell associated with the two sides of it) is given by
\begin{equation}
h^{1,2}_{ij}= \left. g^{1,2}_{\mu\nu}\frac{\partial X^{\mu}_{1,2}}{\partial\xi^{i}}\frac{\partial X^{\nu}_{1,2}}{\partial\xi^{j}}\right| _{\Sigma },
\end{equation}
and the second fundamental form (or extrinsic curvature) of the shell $K_{ij}$ is calculated from
\begin{equation}
K_{ij}^{1,2 }=-n_{\gamma }^{1,2 }\left. \left( \frac{\partial ^{2}X^{\gamma
}_{1,2} } {\partial \xi ^{i}\partial \xi ^{j}}+\Gamma _{\alpha \beta }^{\gamma }
\frac{ \partial X^{\alpha }_{1,2}}{\partial \xi ^{i}}\frac{\partial X^{\beta }_{1,2}}{
\partial \xi ^{j}}\right) \right| _{\Sigma },
\label{sff}
\end{equation}
where
\begin{equation}
n_{\gamma }^{1,2 }=\pm \left\{ \left. \left| g^{\alpha \beta }_{1,2}\frac{\partial \mathcal{G}}{\partial
X^{\alpha }_{1,2}}\frac{\partial \mathcal{G}}{\partial X^{\beta }_{1,2}}\right| ^{-1/2}
\frac{\partial \mathcal{G}}{\partial X^{\gamma }_{1,2}} \right\} \right| _{\Sigma },
\end{equation}
is the unit normal for each side of the shell. In this equation, the upper sign refers to $\mathcal{M}^{2}$ and the lower one to $\mathcal{M}^{1}$. For the metrics given by Eq. (\ref{metric}), they read
\begin{equation}
n_{\gamma }^{1,2 }=\pm \left(-\dot{a},\frac{\sqrt{A_{1,2}(a)+\dot{a}^2}}{A_{1,2}(a)},0,0 \right).
\end{equation}
On $\Sigma $, we can use of the orthonormal basis $\{ e_{\hat{\tau}}=e_{\tau }, e_{\hat{\theta}}=a^{-1}e_{\theta }, e_{\hat{\varphi}}=(a\sin \theta )^{-1} e_{\varphi }\} $ for which the first fundamental form is $h^{1,2}_{\hat{\imath}\hat{\jmath}}= \mathrm{diag}(-1,1,1)$, and the components of the second fundamental form are
\begin{equation} 
K_{\hat{\tau}\hat{\tau}}^{1,2 }=\mp\left(  \frac{\ddot{a}}{\sqrt{A_{1,2} (a) +\dot{a}^2}}+\frac{1}{2}\frac{A'_{1,2}(a)}{\sqrt{A_{1,2}(a) +\dot{a}^2}} \right)
\label{Ktau}
\end{equation}
and
\begin{equation} 
K_{\hat{\theta}\hat{\theta}}^{1,2 }=K_{\hat{\varphi}\hat{\varphi}}^{1,2
}=\pm \frac{1}{a}\sqrt{A_{1,2} (a) +\dot{a}^2},
\label{Kth}
\end{equation}
where the prime represents the derivative with respect to $r$. In this basis, we see that the angular components $K_{\hat{\theta}\hat{\theta}}$ and $K_{\hat{\phi}\hat{\phi}}$ are the same and the junction condition (\ref{jc1}), i.e. $[h_{\hat{\imath}\hat{\jmath}}]=0$, is immediately fulfilled. The jump of the trace of the extrinsic curvature is given by
\begin{equation}
\label{K}
[K]=\frac{2 \left( \sqrt{A_1(a) +  \dot{a} ^2} + \sqrt{A_2(a) +  \dot{a} ^2} \right)}{a}
+ \frac{A_1'(a) + 2 \ddot{a}}{2\sqrt{A_1(a) +  \dot{a} ^2}}
+ \frac{A_2'(a) + 2 \ddot{a}}{2\sqrt{A_2(a) +  \dot{a}^2}} .
\end{equation}
By using the definition of $\Omega$ introduced in Eq. (\ref{Mdef}) and considering that the angular components of the unit normal vanish, we obtain the expression 
\begin{equation}
\label{M-sphe}
\Omega =\frac{\left( A_{2}(a)+2\dot{a}^{2}\right) \phi'_{2}(a)}{\sqrt{A_{2}(a)+\dot{a}^{2}}}+\frac{\left( A_{1}(a)+2\dot{a}^{2}\right) \phi'_{1}(a)}{\sqrt{A_{1}(a)+\dot{a}^{2}}}.
\end{equation}
The junction condition (\ref{jc4}) takes the form
\begin{equation}
-\Omega +4 f'(\phi)\left(\frac{ \sqrt{A_1(a)+ \dot{a}^{2}}+\sqrt{A_2(a)+ \dot{a}^{2}}}{a}+\frac{A_1'(a)+2\ddot{a}}{4\sqrt{A_1(a)+ \dot{a}^{2}}}+\frac{A_2'(a)+2\ddot{a}}{4\sqrt{A_2(a)+ \dot{a}^{2}}}\right)=0,
\label{junCond}
\end{equation}
with $\Omega$ given by Eq. (\ref{M-sphe}). In the orthonormal basis, the energy-momentum tensor for a perfect fluid adopts the form $S_{\hat{\imath}\hat{\jmath} }={\rm diag}(\sigma ,p_{\hat{\theta}},p_{\hat{\varphi}})$, with $\sigma$ the surface energy density and $p=p_{\hat{\theta}}=p_{\hat{\varphi}}$ the transverse pressure. Then, the junction condition given by Eq. (\ref{jc3}) can be written in terms of $\sigma$ and $p$ as 
\begin{equation} 
\sigma=-2 \Omega f'(\phi)-\frac{4 f(\phi)}{a}(\sqrt{A_1(a)+ \dot{a}^{2}}+\sqrt{A_2(a)+ \dot{a}^{2}})
\label{enden}
\end{equation}
and
\begin{equation}
p=2 \Omega f'(\phi)+ f(\phi)\left(\frac{2(\sqrt{A_1(a)+ \dot{a}^{2}}+\sqrt{A_2(a)+ \dot{a}^{2}})}{a}+\frac{A_1'(a)+2\ddot{a}}{\sqrt{A_1(a)+ \dot{a}^{2}}}+\frac{A_2'(a)+2\ddot{a}}{\sqrt{A_2(a)+ \dot{a}^{2}}}\right) ,
\label{pres}
\end{equation}
where $\Omega$ is shown in Eq. (\ref{M-sphe}). All the equations above are greatly simplified if the wormhole spacetime is symmetric across the throat, i.e. when the solutions adopted at both sides of the shell are the same.

\section{Stability of wormholes symmetric across the throat}\label{stab}

For the study of the stability of the static configurations with a throat radius denoted by $a_0$, one usually aims to express $\dot{a}^{2}$ in the form of a master equation
\begin{equation}
\dot{a}^{2}=-V(a),
\label{condicionPot}
\end{equation}
where $V(a)$ can be understood as a potential. The stable solutions under radial perturbations are those that fulfill the condition $V''(a_0)>0$. In our case, it appears quite cumbersome to solve the dynamics in terms of $\dot{a}^{2}$ in a generalized way. However, our study is simplified for thin-shell wormholes symmetric across the throat. In this case, the two spacetimes that are joined at $\Sigma $ are the same, thus  $A_1(r)=A_2(r)=A(r)$, $\phi_1(r)=\phi_2(r)=\phi(r)$, and also $\phi'_1(r)=\phi_2'(r)=\phi'(r)$. Consequently, from Eq. (\ref{M-sphe}) we obtain
\begin{equation}
\Omega =\frac{2\left( A(a)+2\dot{a}^{2}\right) \phi'(a)}{\sqrt{A(a)+\dot{a}^{2}}}.
\end{equation}
The junction condition (\ref{jc4}), which for a spherical shell is shown in Eq. (\ref{junCond}), then takes the form 
\begin{equation}
-\Omega +8f'(\phi)\left(\frac{ \sqrt{A(a)+ \dot{a}^{2}}}{a}+\frac{A'(a)+2\ddot{a}}{4\sqrt{A(a)+ \dot{a}^{2}}}\right)=0. 
\label{JC}
\end{equation}
We can rewrite it as
\begin{equation}
- a \left( A(a)+2\dot{a}^{2}\right) \phi'(a)+ 4f'(\phi)\left( A(a)+ \dot{a}^{2}\right) +af'(\phi)\left( A'(a)+2\ddot{a}\right) = 0 ,
\label{JCsimpler}
\end{equation}
or reordering terms
\begin{equation}
2af'(\phi)\ddot{a} + \left( 4f'(\phi) - 2a \phi'(a)\right) \dot{a}^{2}+af'(\phi)A'(a)+ \left(  4f'(\phi)- a \phi'(a)\right) A(a) = 0 .
\label{JCsimpler2}
\end{equation}
We look for an expression of $\dot{a}^2$ in the form of Eq. (\ref{condicionPot}). As we are considering a non-minimally coupled scalar field, we assume that $f'(\phi)\ne 0$. With the help of the definitions
\begin{equation}
z(a)=\dot{a}^{2}+A(a) , \qquad  y(a)=\frac{4}{a} - \frac{2\phi'(a)}{f'(\phi )} , \qquad  x(a)= \frac{A(a)\phi'(a)}{f'(\phi )}  ,
\end{equation}
and using that $2\ddot{a}=d\dot{a}^2/da$, we can rewrite Eq. (\ref{JCsimpler2}) in the form 
\begin{equation}
z'(a)+y(a)z(a)+x(a)=0 ,
\label{difEq}
\end{equation}  
which is a first order ordinary linear equation with non-constant coefficients. For regular enough functions $x(a)$ and $y(a)$ this equation can be formally solved to give $z(a)$, and consequently it allows to obtain $\dot{a}^2= z(a)-A(a)\equiv -V(a)$. But from a practical point of view, in the particular cases, as $x(a)$ and $y(a)$ depend on the metric function $A(a)$ and also on the functions $f(\phi)$ and $\phi(a)$ associated to the scalar field, finding the expression of the potential and its first and second derivatives can be a rather difficult task.

An interesting situation where the potential can be readily calculated is when $\phi'(a)=0$, for which we have that $[n^{\mu} \partial_{\mu}\phi]=0$ and then $[K]=0$. In this case, from Eq. (\ref{JCsimpler}) the throat radius should fulfill
\begin{equation}
4f'(\phi)\left( A(a)+ \dot{a}^{2}\right) +af'(\phi)\left( A'(a)+2\ddot{a}\right) = 0 ,
\label{JCsimpler3}
\end{equation}
which for the static configuration reduces to
\begin{equation}
4A(a_0)+a_0A'(a_0)=0
\label{JCstatic}
\end{equation}
when $f'(\phi)\neq 0$. The differential equation (\ref{difEq}) takes the simple form
\begin{equation}
z'(a)+ \frac{4}{a} z(a)=0 ,
\label{difEqSimp}
\end{equation}  
having the solution $z(a)=C/a^4$, with $C$ a constant. The potential then reads
\begin{equation}
V(a)=A(a)-\frac{a_0^4 }{a^4}A(a_0),
\label{pot}
\end{equation}
where the constant $C$ has been obtained from the condition $V(a_0)=0$. The second order Taylor expansion of this potential takes the form
\begin{equation}
V(a)=V(a_0)+V'(a_0)(a-a_0)+\frac{V''(a_0)}{2}(a-a_0)^2+\mathcal{O}(a-a_0)^3 .
\end{equation}
Besides from $V(a_0)=0$, with the aid of Eq. (\ref{JCstatic}) it is not difficult to verify that $V'(a_0)=0$, while the second derivative of the potential evaluated at $a_0$    reads
\begin{equation}
\label{stability}
V''(a_0)= A''(a_0) -20 \frac{A(a_0)}{a_0^2}.
\end{equation}
The stable solutions satisfy that $V''(a_0)>0$. 

The perfect fluid at the shell has an energy density (\ref{enden}) and a pressure (\ref{pres}), whose expressions in the static case read, respectively
\begin{equation} 
\sigma_0=-\frac{8 f(\phi_0)}{a_0}\sqrt{A(a_0)}
\label{enden0}
\end{equation}
and
\begin{equation}
p_0=2 f(\phi_0)\left(\frac{2\sqrt{A(a_0)}}{a_0}+\frac{A'(a_0)}{\sqrt{A(a_0)}}\right),
\label{pres0}
\end{equation}
where $\phi_0=\phi(a_0)$. By using Eq. (\ref{JCstatic}) we can easily see that
\begin{equation}
p_0=-\frac{4 f(\phi_0)}{a_0}\sqrt{A(a_0)},
\label{pres1}
\end{equation}
so the equation of state is fixed, having the form $p_0=\sigma _0 /2$. At this point, it is worthy to emphasize that the relation between the jump of the trace of the second fundamental form and the scalar field given by Eq. (\ref{jc4}), which for this class of wormholes results in $[K]=0$, leads to the constraint on the shell radius $a_0$ shown in Eq. (\ref{JCstatic}). This constraint, not present in GR, also fixes the equation of state at $\Sigma $ in the presence of the scalar field.

Another important aspect to analyze is the type of matter that constitutes the wormhole, which can be studied in terms of the weak energy condition (WEC). This condition, given by the inequalities
\begin{equation}
\label{WEC}
\sigma_0\ge 0 \qquad \mathrm{and} \qquad \sigma_0 +p_0 \ge 0,
\end{equation} 
determines when the matter at the throat satisfies WEC (dubbed normal) or not (exotic). The non-minimal coupling function $f(\phi )$ has to be positive everywhere in order that the effective gravitational constant satisfy $G_{eff}\propto 1/f(\phi_0)>0$ and (from the quantum point of view) avoid the graviton to be a ghost \cite{Bronnikov:2009tv}. Considering Eq. (\ref{enden0}) and adopting $f(\phi_0)>0$, we realize that the energy density at the throat is always a negative quantity and this implies the existence of exotic matter for any choice of the parameters. This fact has actually been proved for any symmetric thin-shell wormhole configuration in any scalar-tensor theory \cite{Bronnikov:2009tv}. In other terms, the price to pay for having matter satisfying WEC at the throat, which requires $f(\phi_0)<0$, is to have $G_{eff}<0$ and a ghost graviton.

\section{Wormholes with a radial electric field }\label{scalartensor}

In this section, we consider a particular case of Einstein-Maxwell theory coupled to a conformally invariant scalar field. We adopt the solution found in \cite{Astorino:2013sfa} to construct the thin-shell wormholes and we analyze the stability of the static configurations and the fulfillment of the WEC.

The action for Einstein-Maxwell theory with a conformally coupled self interacting scalar field $\phi$ introduced in \cite{Astorino:2013sfa} is given by Eq. (\ref{action1}) with
\begin{equation}
\label{f(Phi)}
f(\phi)=\frac{1}{16\pi G}-\frac{\phi^2}{12},
\end{equation}
\begin{equation}
U(\phi)=0,
\end{equation}
\begin{equation}
\label{LEM}
\mathcal{L}_{M} =-\frac{1}{16\pi}F_{\mu\nu}F^{\mu\nu}.
\end{equation}
where $F_{\mu \nu }=\partial _{\mu }\mathcal{A}_{\nu } -\partial _{\nu }\mathcal{A}_{\mu }$ is the electromagnetic tensor in terms of the vector potential $\mathcal{A}_{\mu}$. The corresponding gravitational, scalar, and electromagnetic field equations are respectively
\begin{equation}
\label{EFEm1}
G_{\mu\nu}=8\pi G \left( T^{S}_{\mu\nu}+T^{EM}_{\mu\nu}\right) ,
\end{equation}
\begin{equation}
\label{EFEf1}
\Box\phi=\frac{1}{6}R\phi,
\end{equation}
\begin{equation}
\label{EFEem1}
\partial_{\mu}(\sqrt{-g}F^{\mu\nu})=0,
\end{equation}
where the sum of
\begin{equation}
\label{TS}
T^{S}_{\mu\nu}=\partial_{\mu}\phi\partial_{\nu}\phi-\frac{1}{2}g_{\mu\nu}\partial_{\sigma}\phi\partial^{\sigma}\phi+\frac{1}{6}(g_{\mu\nu}\Box-\nabla_{\mu}\nabla_{\nu}+G_{\mu\nu})\phi^{2},
\end{equation}
and
\begin{equation}
\label{TEM}
T^{EM}_{\mu\nu}=\frac{1}{\mu_{0}}(F_{\mu\rho}F_{\nu}^{\;\rho}-\frac{1}{4}g_{\mu\nu}F_{\rho\sigma}F^{\rho\sigma})
\end{equation}
is the total energy-momentum tensor $T_{\mu\nu}$.

For the construction of the thin-shell wormholes, we are interested in the static hairy Reissner-Nordstr\"{o}m black hole, which is a solution of the field equations with a conformally coupled constant scalar field. In this case, adopting units so that $G=1$ and $\mu_0/ (4\pi ) =1$, the only non-zero component of the electromagnetic field that solves (\ref{EFEem1}) can be expressed as
\begin{equation}
\label{EMF}
\partial_{r}A_{t}=F_{rt}=\frac{Q}{r^{2}},
\end{equation} 
and the metric and scalar field which are solutions of (\ref{EFEm1}) and  (\ref{EFEf1}), respectively read \cite{Astorino:2013sfa}
\begin{equation}
ds^2 = -\left(1-\frac{2m}{r}+\frac{Q^2+s}{r^2}\right)dt^2+\left(1-\frac{2m}{r}+\frac{Q^2+s}{r^2}\right)^{-1}dr^2+r^2(d\theta^2+ \sin^2\theta d\phi^2)\, 
\end{equation}
\begin{equation} 
\label{scalar}
\phi=\pm\sqrt{\frac{6}{8\pi } }\sqrt{\frac{s}{s+Q^2}},
\end{equation} 
where $m$ is the mass, $Q$ is the electric charge, and $s$ the scalar hair. The sign of $Q$ does not affect the metric nor the scalar field (it only has influence on the direction of the radial electric field), so we adopt $Q \ge 0$ without losing generality. Because of Eq. (\ref{scalar}), for positive values of $s$, we see that $Q$ can take any value, while for negative values of $s$, we should have $Q^2<-s$. For $s=0$, there is a null scalar field and we recover the Reissner-Nordstr\"{o}m solution of GR, in which $Q$ has no restrictions. The horizons are given by
\begin{equation}
r_{\pm }=m\pm\sqrt{m^2 - Q^2 - s} ,
\end{equation}
when $m^{2}- Q^2 - s \ge 0$. The plus sign corresponds to the event horizon $r_{h}$ and the minus sign to a Cauchy horizon. For $m^{2}- Q^2 - s<0$  no horizons are present and there is a naked singularity at the origin, because the Kretschmann invariant $\mathcal{K}=R_{\alpha\beta\gamma\delta} R^{\alpha\beta\gamma\delta}$ (quadratic in the Riemann tensor)
\begin{equation}
\mathcal{K} = \frac{48(Q^{2}-mr+s)^{2}}{r^{8}}
\end{equation}
diverges. When $m^{2}- Q^2 - s = 0$ the two horizons coincide (extreme case). Note that a necessary condition for the presence of the horizons is that $m^{2}\ge s$; in this case there is a critical value of the charge $Q_c$, which reads
\begin{equation}
Q_c=\sqrt{m^2 - s}
\end{equation}
so that there are two horizons for $Q< Q_c$, one horizon when $Q= Q_c$, and no horizons if $Q >Q_c$. 

\begin{figure}[t!]
 \centering
 \includegraphics[width=1\textwidth]{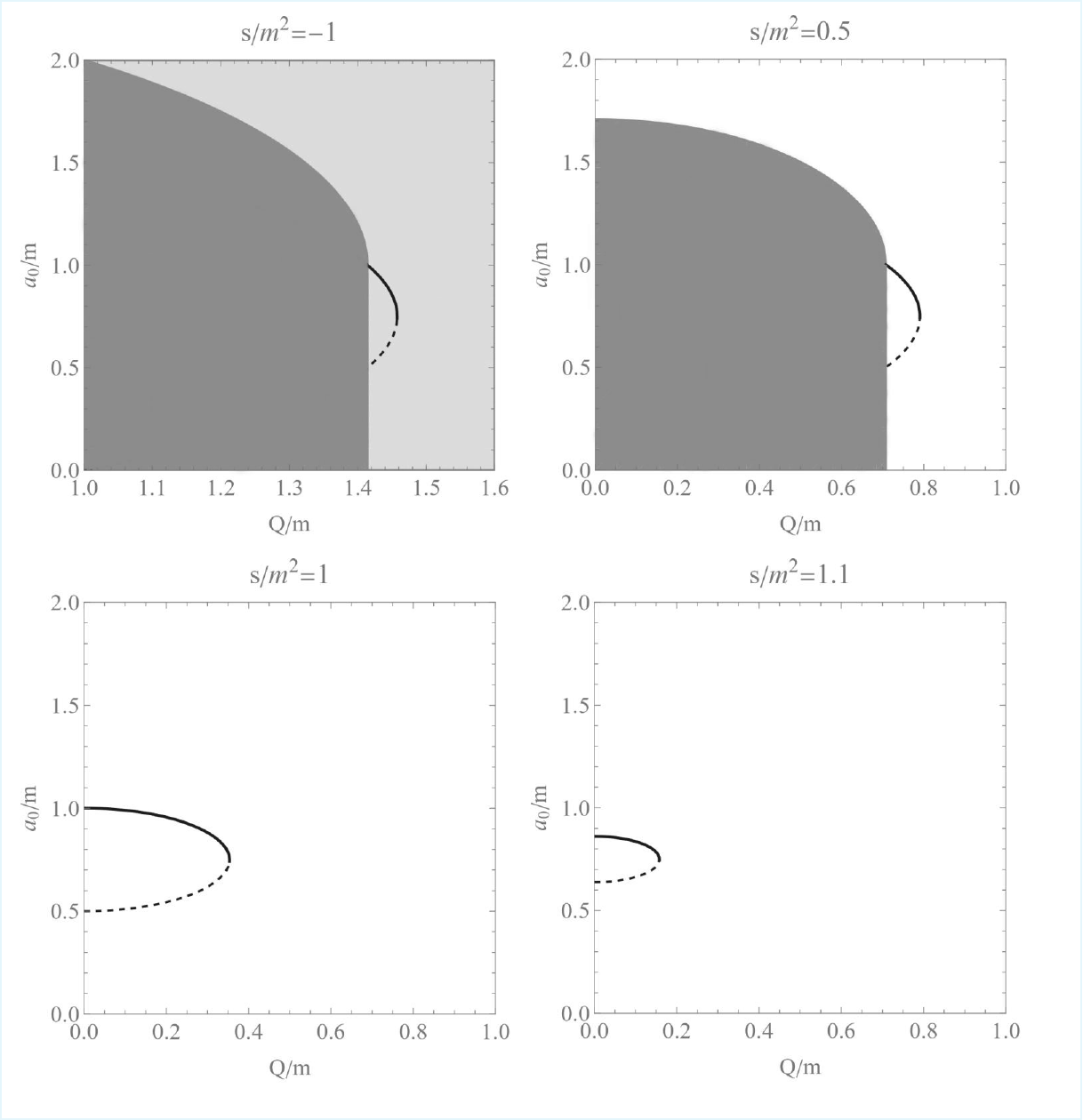}
 \caption{Stability of a spherical thin-shell wormholes symmetric across the throat, in Einstein-Maxwell theory with a conformally coupled and constant scalar field, for different values of the hair parameter $s/m^2$, with $m$ the mass. The solid line represents the stable static configurations with radius $a_0/m$ and charge $Q/m$, while the dashed line shows the unstable solutions. The dark gray regions have no physical meaning (see text) and in the light gray ones the scalar field is not real.}
 \label{fig:grid}
\end{figure}

In order to obtain a traversable wormhole, the throat radius should satisfy $a_0 > r_{h}$ when $Q \le Q_c$, which removes the horizon and the singularity at the origin,  and $a_0>0$ if  $Q > Q_c$, which takes away the naked singularity at the origin. The scalar field shown in Eq. (\ref{scalar}) satisfies that $\phi'(r) =0$, then we use the formalism of the previous section to analyze the behavior of the static configurations, with the potential given by Eq.  (\ref{pot}). The possible values of $a_0$ are those which satisfy Eq. (\ref{JCstatic}) and the stability is determined by Eq. (\ref{stability}). The scalar field should be real, which restricts the value of charge to $Q^2<-s$ when $s<0$. 

We present our results graphically. In Fig. \ref{fig:grid}, we study the existence and stability of thin-shell wormholes by displaying $a_0/m$ against $Q/m$ for different values of the parameter $s/m^2$. The black curve corresponds to the solutions that satisfy (\ref{JCstatic}), where the stable configurations are drawn with a solid line and the unstable ones with a dashed line. The dark gray areas show the regions without physical meaning which are removed, while the light gray ones are  those where the scalar field is not real. In more detail, for negative $s/m^2$, represented by the plot with $s/m^2=-1$, we find two solutions for $Q>Q_c$, the larger one is stable, while the other is unstable, but in both the scalar field is not real so they have to be discarded. For $0<s/m^2<1$, represented by $s/m^2=0.5$, two solutions are found when $Q>Q_c$, a stable, larger one and an unstable, smaller one. The region without physical meaning shrinks to finally disappear as we increase the value of $s/m^2$. For $s/m^2=1$, there are two solutions for small values of $Q/m^2$; the larger one is stable while the other is unstable. In this case, there are no horizons in the original metric for $Q \neq 0$, and for the case of $Q=0$, there is a horizon where the solutions fall on its border or into it. For $1<s/m^2 < 9/8$, represented by $s/m^2=1.1$ the behavior is similar to the case with $s/m^2=1$. Finally, if $s/m^2\ge 9/8$, one can see that Eq. (\ref{JCstatic}) has no real solutions, so our construction is not possible.

Let us analyze the matter content at the throat. When $f(\phi_0)>0$, that is $s/(s+Q^2) < 1$, the WEC is not fulfilled since $\sigma _0 < 0$ in Eq. (\ref{enden0}). If $f(\phi_0) \le 0$, that is $s/(s+Q^2) \ge 1$, the WEC is satisfied, because $\sigma _0\ge 0$ and $\sigma _0 +p_0 \ge 0$, with $p_0$ given by Eq. (\ref{pres1}). When $s>0$ we can see that the matter is always exotic,  while for $s<0$, if $Q^2<-s$ it is normal and if $-s< Q^2$ is exotic but with a non-real scalar field. Outside the shell, i.e. in the bulk, the total energy-momentum in an orthonormal frame reads \cite{Astorino:2013sfa}
\begin{equation}
\label{emt-bulk1}
T^{\hat{\mu }\hat{\nu }} =\frac{Q^2+s}{r^4}\mathrm{diag}\left( 1,-1,1,1 \right),
\end{equation} 
which for a perfect fluid takes the form $T^{\hat{\mu }\hat{\nu }} =\mathrm{diag}(\rho ,p_{1},p_{2},p_{3})$, with $\rho $ the volume energy density and $p_i$ the pressures.  It is easy to check that $\rho \ge 0$ and $\rho + p_{i} \ge 0 $ if $Q^2+s \ge 0$, so the WEC is satisfied at the bulk in this case. In particular, when $s>0$ the matter is always normal, while for $s<0$  if $Q^2 \le -s$ it is exotic and if $-s \le Q^2$ is normal but with a non-real scalar field.

In brief, we have found that the configurations with $s<0$ are not physically relevant and those with $s>0$ present interesting results. In particular, if $0< s/m^2 <1$ there is a pair of solutions with large values of $Q/m$, i.e. $Q> Q_c$, one stable and the other unstable, while for $1\le s < 9/8$ the stable and the unstable configurations correspond to small values of $Q/m$. In both scenarios, the WEC is satisfied at the bulk, but not at the throat.

\section{Conclusions}\label{conclu}

In this work, we have constructed a class of spherical thin-shell wormholes in scalar-tensor theories with a self-interaction potential non-minimally coupled to gravity, in which the throat is located at the shell. For wormholes symmetric across the throat, we have studied the stability of the static configurations under perturbations preserving the spherical symmetry, establishing a condition in terms of the second derivative of a potential.

We have found that when $[n^{\mu}\partial_{\mu}\phi]= 0$, which implies $[K]= 0$, the equations are greatly simplified and resemble those in the $F(R)$ gravity counterpart. The condition $[K]=0$ always holds in the junction formalism of $F(R)$ theories \cite{Senovilla:2013vra}. In our case, the expressions for the condition satisfied by the throat radius $a$, the energy density $\sigma$, and the pressure $p$, have a similar form to the corresponding ones for thin-shell wormholes in $F(R)$ theories. They differ by a factor proportional to $f(\phi)$ instead of a function of the Ricci scalar $R$. This behavior is not surprising, since  $F(R)$ gravity and scalar-tensor theories are closely related. However, the solutions to use in the wormhole construction depend on the theory adopted, so the specific models present different characteristics.

In particular, we have considered the example of charged thin-shell wormholes with spherical symmetry, by using known black hole solutions \cite{Astorino:2013sfa} in Einstein-Maxwell theory with a conformally self-interacting scalar field $\phi (r)$. We have studied the stability of the static configurations under radial perturbations and analyzed the fulfillment of the WEC at the throat and in the bulk. The Lagrangian of the theory is quadratic in the scalar field, the spacetime adopted has a constant $\phi $ and a metric depending on the mass $m$, the charge $Q$, and the scalar hair $s$. We have concluded that there is a pair of solutions, one stable and the other unstable, for a range of values of $s/m^2$ and $Q/m$. The stable configuration appears when $0<s/m^2 <1$ for large $Q>Q_c$ and when $1 \le s/m^2 < 9/8$ for comparatively small values of $Q/m$; in both cases with $a_0/m$ close to $1$. We have verified that the existence of exotic matter at the throat cannot be avoided when requiring the physically sound condition that $f(\phi )>0$, since for a wormhole symmetric across the throat the energy conditions are always violated in any scalar-tensor theory, as it has been already proved \cite{Bronnikov:2009tv}.

For future research, it would be interesting to consider in detail more complex scenarios in which $[n^{\mu}\partial_{\mu}\phi] \neq 0$, especially to study the stability and to analyze whether the presence of exotic matter can be eliminated or reduced.

\section*{Acknowledgments}

EFE and GFA acknowledge the support by CONICET. VK thanks IAFE for the hospitality and acknowledges the financial support from Erasmus+ EU funds and the national co-financing grant from Estonian Ministry of Education and Science for Erasmus+ mobility activity.

\end{document}